\documentclass[11pt,a4paper]{article}
\usepackage[hyperref]{emnlp2020}
\usepackage{times}
\usepackage{latexsym}

\usepackage{microtype}

\aclfinalcopy 

\title{Examining the Feasibility of Off-the-Shelf Algorithms for Masking Directly Identifiable Information in Social Media Data}

\author{Rachel Dorn,\textsuperscript{1} Alicia L. Nobles,\textsuperscript{2} Masoud Rouhizadeh,\textsuperscript{3} Mark Dredze\textsuperscript{1} \\
        \textsuperscript{1}Department of Computer Science, Johns Hopkins University \\
        \textsuperscript{2}Department of Medicine, University of California San Diego\\
        \textsuperscript{3}Institute for Clinical \& Translational Research, Johns Hopkins University \\
        \normalsize{rachel.m.dorn@gmail.com, alnobles@health.ucsd.edu, mrou@jhu.edu, mdredze@cs.jhu.edu} 
  }

\date{}

\begin{document}
\maketitle
\begin{abstract}
The identification and removal/replacement of protected information from social media data is an understudied problem, despite being desirable from an ethical and legal perspective. This paper identifies types of potentially directly identifiable information (inspired by protected health information in clinical texts) contained in tweets that may be readily removed using off-the-shelf algorithms,  introduces an English dataset of tweets annotated for identifiable information, and compiles these off-the-shelf algorithms into a tool (Nightjar) to evaluate the feasibility of using Nightjar to remove directly identifiable information from the tweets. Nightjar as well as the annotated data can be retrieved from \url{https://bitbucket.org/mdredze/nightjar}.
\end{abstract}

\section{Introduction}
Social media data are increasingly used in many research areas \cite{zimmer2014topology,weller2014twitter}, including medicine and public health \cite{paul2017social,sinnenberg2017twitter}, because of its accessibility and volume. For example, social media data has been leveraged to develop models that detect mental health states \cite{Mowery,Coppersmith,Choudhury}.

The influx of social media data in research, especially within medicine and public health, has shifted discourse to consider how best to protect the privacy of the users who generated the data. In contrast to the emergent field of social media research, medical research has established stringent policies to protect patient privacy including the removal of protected health information (PHI) in clinical text used for research. For example, the US Health Insurance Portability and Accountability Act of 1996 (HIPAA) established 18 categories of PHI that should be removed from clinical text used in research \cite{HIPAA}. Many of these categories, like social security numbers, are unlikely to be present in social media data.

The environment around social media research is dramatically different. While there are proposed protocols for protection of user data \cite{Benton}, limited work has examined what types of identifiable information may be present in social media data that needs to be protected and how best to remove such information \cite{Nguyen-Son_2012,Beigi_2019}. Additionally, even with removal of such information, one may still be able to reverse identify a social media user potentially exposing users to unanticipated risk \cite{Kelley_conductingresearch,loosetweets}. 

In practice, few health-related social media studies and dataset releases remove potentially identifiable information despite the widespread recognition of the sensitivity of this research \cite{conway2014ethical}. An example of an exception is the CLPsych 2015 Shared Task \cite{CLPsych}, which attempted to de-identify Twitter data by removing metadata (e.g., geolocation) and salting usernames and URLs. However, the task did not examine what sensitive information may still be present in the tweets.

In this study, we (1) discuss types of potentially directly identifiable information present in Twitter data that may be readily removed using publicly available or easily implementable algorithms, (2) annotate a Twitter dataset for these identifiable information, and (3) evaluate the performance of these combined algorithms for removal of the identifiable information on our annotated dataset. We release our annotated Twitter dataset as well as Nightjar, our system that combines these algorithms into an easily implementable Python tool, to build on.

\begin{table}[t]
\begin{small}
\begin{tabular}{ll}
Names & Health plan beneficiary numbers \\
Dates, except year & Certificate/license numbers \\
Telephone numbers & VIN and license plates \\
Geographic data & Web URLs \\
FAX numbers & Device identification numbers \\
Social Security numbers & IP addresses \\
Email addresses & Full face photos \\
Medical record numbers & Biometric identifiers \\
Account numbers & Unique identification numbers \\
\end{tabular}
\end{small}
\caption{\label{tab:PHI} HIPAA Protected Health Information}\smallskip
\end{table}

\section{Background}
\textbf{Identifiable Information in Social Media.}
Although most social media research uses publicly available social media data, research, especially on sensitive topics like health, could expose users to risk. For example, while someone may publicly share a mental health diagnosis on Twitter \cite{coppersmith2014quantifying}, they may be exposed to risk if their username is included in a publication or released dataset. Many social media users are largely unaware that their data may be used in research and intend for their postings to only be seen by a limited audience \cite{Fiesler}. Additionally, although many users are supportive of their social media being used for research, they think data about sensitive issues, such as mental health, should be protected similar to how clinical text is protected under HIPPA \cite{Andalibi_2020}. 

Some research has focused on privacy preserving techniques for user-generated data. For example, \cite{Beigi_2019} focused on a privacy preserving representation learning framework for user-generated data that preserved semantic meaning of the original text. However, there is no particular emphasis placed on identifying what types of information present in the data may need to be protected. With limited research available on the types of identifiable information that may be present in social media data and methods to remove such information, we turn to the well established literature on removal of PHI from clinical text.

\textbf{PHI in Clinical Text.} Under HIPPA, there are 18 types of PHI in clinical text (see Table~\ref{tab:PHI}) that must be de-identified for downstream usages. Years of research have developed automated systems, including the 2006 and 2014 i2b2 shared tasks \cite{stubbs2015automated}. De-identification of clinical text is (mostly) akin to named entity recognition (NER): find the named entities and remove/replace them. The most successful systems utilize hybrid approaches \cite{Yang}---combining rule-based and machine-learning methods---or, in recent years, deep learning \cite{Dernoncourt}. Removal strategies are dependent on downstream usage. For example, a dataset constructed for automated analysis within the same institution will have less need to de-identify, whereas a publicly released dataset must ensure all PHI is removed. Additionally, these de-identification approaches aim to retain statistical properties of the data to ensure that the de-identified clinical text does not degrade the effectiveness of the NLP tool.

\section{Methodology}
We describe the types of potentially directly identifiable information, partly motivated by the PHI defined by HIPAA, that we hypothesize can be removed from the tweet text using publicly available or easily implementable algorithms below. We combine these algorithms into {\tt Nightjar}\footnote{Nightjars are birds known for camouflage since they have soft plumage and variegated coloring that blends into the surrounding.}, a Python tool that uses off-the-shelf algorithms to automate removal of this information in tweets.\footnote{\url{https://bitbucket.org/mdredze/nightjar}}

\textbf{URLs.} 
Although most tweeted URLs do not contain identifiable information (e.g., news in comparison to a personal website), ensuring a website is indeed not personally identifiable would require downloading and analyzing the linked content. Therefore, Nightjar simply removes URLs from the tweets using regex, which should have minimal affect on research since most downstream analyses do not attend to the content of the URL string.

\textbf{Usernames.} 
Mentions of usernames (@TwitterUser) are the social media equivalent of names and unique identification numbers in PHI. Depending on their usage, they can be used to identify the user.

\textbf{Phone Numbers.}
While many tweeted phone numbers may be unrelated to the user, in some cases they could be linked back to a individual person. Additionally, phone numbers are not likely to impact downstream research. Nightjar removes phone numbers from a tweet using regex if the account is not verified. 

\textbf{Email Addresses.}
Similarly, tweeted email addresses are often not related to the user, but in some cases could be linked back to an individual person. Nightjar removes email addresses from a tweet using regex if the account is not verified. 

\textbf{Identification Numbers.}
PHI includes several types of identification numbers (Table~\ref{tab:PHI}) like social security numbers and vehicle identification numbers. Although these are not likely to be tweeted, Nightjar incorporates functionality to remove long alphanumeric strings using regex.

\textbf{Names of People, Organizations, Groups, and Locations.}
The aforementioned potentially directly identifiable information is easily identified and removed using regex. Other information types are more challenging to identify - names of people, organizations, groups, and locations. We turn to publicly available NER algorithms. We combine two NER algorithms, the English models of CoreNLP \cite{manning} and spaCy \cite{spacy}, to recognize names with greater recall. If either model marks a token as identifiable information, Nightjar removes the token and implements a synthetic replacement strategy for the identified entity. It replaces mentions of these entities with random entries of the same type. For example, consider a person mentioned in a tweet. If the original tweet states ``\textit{Shout out to Katie for making this event happen},'' Nightjar will replace ``\textit{Katie}'' with a random name of a person changing the tweet to ``\textit{Shout out to Brian for making this event happen}.'' Replacement values for entities based on entity type are generated by Faker. 

\section{Evaluation}
\textbf{Dataset and Annotation.}
We sampled 2000 tweets on February 2, 2018 from the 1\% Twitter streaming API and filtered to English only tweets (identified by a metadata label) resulting in a total of 709 tweets for annotation. 

The first author labeled phone numbers, email addresses, identification numbers, persons, organizations, groups, and locations at the token level for each tweet. Only people who were not public figures were annotated as persons. Deciding who is a public figure is not an easy task. While national public figures are easily recognizable, local public figures, like the name of a mayor, are less so. Organizations were defined as public entities such as company names and groups were defined as nationalities, religions, and other affiliations. Because of the difficulty of labeling public figures and organizations, we relied on Twitter metadata to indicate if a person or organization was verified (i.e., Twitter considers the person or organization of ``public interest''). Locations included zip code, city, state, and country - all entities detected by the NER approaches employed in Nightjar. Table \ref{tab:phi_results} presents the resulting frequencies of each label in the 709 tweets.

\textbf{Performance of Nightjar.}
We evaluated Nightjar's performance on discovery of these annotated labels using precision, recall, F1, and all-or-nothing recall, the latter being a more conservative measure of discovery because it only credits algorithms only that find all instances of annotated labels in a particular tweet \cite{Scaiano}.

\section{Results}
\begin{table}[t]
\begin{small}
\begin{tabular}{l|c|c|c|c|c}
{\bf Info Type} & {\bf No} & {\bf P} & {\bf R} & {\bf F1} & {\bf AON R}\\
\hline
URL & 371 & 1.0 & 1.0 & 1.0 & 1.0\\
Username & 368 & 0.685 & 0.991 & 0.81 & 0.989\\
Person & 58 & 0.144 & 0.735 & 0.24 & 0.741\\
Org & 7 & 0.02 & 0.308 & 0.038 & 0.429\\
Group & 2 & 0.0 & 0.0 & 0.0 & 0.0\\
City & 3 & 0.154 & 0.5 & 0.235 & 0.667\\
State & 5 & 0.278 & 1.0 & 0.435 & 1.0\\
Country & 3 & 0.2 & 1.0 & 0.333 & 1.0\\
Location & 1 & 0.037 & 0.333 & 0.067 & 1.0\\
Phone \# & 2 & 1.0 & 1.0 & 1.0 & 1.0\\
\hline
\textbf{Micro Avg} & \textbf{828} & \textbf{0.54} & \textbf{0.961} & \textbf{0.692} & \textbf{0.959} \\
\textbf{Macro Avg} & \textbf{828} & \textbf{0.352} & \textbf{0.687} & \textbf{0.432} & \textbf{0.774} \\
\end{tabular}
\end{small}
\caption{\label{tab:phi_results} Performance of Nightjar on 709 annotated tweets. Identifiable information with no labeled instances in the data are not shown. Their counts in the unlabeled data were: zip code: 61; email address: 13; ID: 0. No = number of labels, P = precision, R = recall, AON = all-or-nothing.}
\end{table}

Table \ref{tab:phi_results} presents the types of potentially directly identifiable information in our annotated dataset and the performance of Nightjar on each type of information. Some identifiable information common to clinical data, including email addresses, zip codes, facility names, and identification numbers, were completely absent from our annotated dataset. These types of information account for nearly half of the 18 PHI categories (see Table \ref{tab:PHI}). Out of our annotated sample, there were 23 false negatives (i.e., missed labels) compared to 298 false positives (i.e., wrongly identified as identifiable information).  

In our annotated dataset, URLs and usernames (from unverified accounts) were the most common types of potentially directly identifiable information present in tweets. Nightjar achieves perfect performance on removal of URLs and reasonable performance on removal of usernames. Phone numbers also occur in the dataset, albeit at much lower frequency (only two instances), and Nightjar achieves perfect performance on these instances. 

In comparison to URLs and usernames, entity-based information types are less common in tweets. Mentions of a person are the most common entity present in tweets. Identification of person entities has lower precision mostly because the combined NER approach falsely identifies a person entity or identifies an entity when one was not present in a tweet. For example, 16 of the 23 false negatives had non-standard capitalization or punctuation, a challenge for NER. 

\section{Discussion}
Overall, Nightjar does a perfect job at identifying URLs and phone numbers, and a reasonable job at recalling usernames and some entities (person, organization, city, state, and country), but it identifies too many false positives resulting in an unacceptably low precision. This is unsurprising because human annotators can read the entire tweet and contextualize if the information contained in the tweet is potentially sensitive, whereas the publicly available NER approaches we evaluated for the task simply identify whether an entity is present in a tweet. Improving recall of NER models is an ongoing challenge in the NLP community. With the current state of performance, using out-of-the-box NER approaches, like we have presented in this study, may pose the risk of degrading the quality of the data (by removing non-sensitive data) making it potentially less useful for downstream research. Our results indicate that the ability to use NER for identification of directly identifiable information should also focus on improving contextualization based on the tweet holistically much like a human annotator can. Recent advances in NLP with deep transformer based encoders can build representations of text that incorporate long span context \cite{devlin-etal-2019-bert,nguyen-etal-2020-bertweet}.
Additionally, entity linking for social media data \cite{Shen_2013,Derdzynski_2015} to a relevant knowledge base could indicate which mentions were to well known, popular entities that need not be masked.

The fact that some identifiable information that are common to clinical text are not tweeted is unsurprising given that streamed Twitter data is public. However, this highlights the need for additional consideration of what may be information that should be protected in social media data compared to how we currently define and protect information in clinical text. For example, specific identifiers (e.g., holding an identity of a vulnerable group) combined with a location and job title may be enough to reverse identify an individual highlighting the dramatically different landscape of social media data. This difference heightens the need for (1) re-considering what is identifying information for social media data and (2) the usefulness of establishing a large dataset focused on removing and de-identifying social media data as well as more robust cross-platform analyses.

Our study indicates several areas for future work. First, we demonstrate that the problem landscape is very different than de-identification of medical documents. The most common identifiable information in tweets (usernames and URLs) are almost entirely absent from clinical notes. HIPAA regulations focus primarily on identification numbers, which are rarely found in social media messages. The most common and challenging identifiable information types in social media rely on named entities, and thus can be potentially identified with NER systems that focus specifically on this task.

Second, unlike in clinical text, removal of identifiable information for social media depends on the context of the information. Person names may or may not identify users depending on the specific name and the context in which they are used. While removal of all names is a reasonable solution, such an approach may limit the usefulness of the dataset in downstream tasks that benefit from knowing the identity of public figures or organizations.

An important limitation of our work is we only consider one social media platform. Other platforms use different username schema, and so removal of one of our most common directly identifiable information types (usernames) may pose a differing challenge. Another limitation is that we only consider the text of the tweet. Tweets can contain images, videos, and, most recently, audio samples, which present their own identification challenges not considered in this study. Additionally, tweets inherently have related metadata, which our study did not consider outside of examining to see if the user was a verified account. 

To support further research on this topic we release Nightjar and our annotated Twitter dataset. We note that while we release an identifiable set of tweets, we believe the risk is minimal since the set was randomly selected and does not reflect a specific sensitive context. Nevertheless, use of our data requires signing a data use agreement that ensures proper use of the data.

Finally, we emphasize that Nightjar did not achieve good performance at removing potentially directly identifiable information from tweets indicating that off-the-shelf algorithms, at the moment, do not achieve reasonable performance.

Future work that improves the underlying technology can lead to more effective masking of identifiable information. Additionally, further study of Twitter data, and other social media platforms, could identify types of information that should be protected beyond those that we have surfaced in this work, which has focused on HIPAA type identifiers. Finally, given how social media data is used in research and distributed, these questions should be asked in the context of a wide range of research efforts. Different types of work could expose users to different risks, impose different requirements on data quality, and have sensitivities to different issues.

\bibliographystyle{acl_natbib}
\bibliography{emnlp2020}

\end{document}